# A barrier/seed system for electroless metallization on complex surfaces using (aminomethylaminoethyl)phenethyltrimethoxysilane self-assembled films

J. K .L. Peters, G. D. Ashby, H. D. Hallen[*]

*Department of Physics, North Carolina State University, Raleigh, North Carolina, 27695, USA*

High frequency signals propagate along the edges of conductors. If the conductors are electroplated, then the seed layer forms at least one edge, so care must be taken to insure the electrical quality of these layers. In this work, we study the initial quality of SAM-based seed layers that are compatible with complex surfaces including through-silicon vias (TSVs), as are used in via-last three-dimensional semiconductor device packaging. For via holes operating at high frequency and low loss, the conformal and electrical quality of the seed metal becomes very important. Other properties of a multifunction seed layer are its abilities as a barrier layer, which protects the substrate from high temperature diffusion of the deposited metal. The barrier layer must be robust enough to withstand diffusion, yet thin enough to provide a conformal surface that allows for metal seed layer deposition. Standard barrier layer deposition methods such as evaporation or sputtering require either a line of sight from the source or aspect ratios large enough to provide scattering from the background gas within the structure to coat all surfaces. Electrochemical and chemical vapor deposition provide alternatives, but concerns arise about contamination and compatibility with radio frequency or high-speed digital signals. We propose a barrier layer based on an aromatic self-assembled monolayer (SAM) for use in electroless copper seed layer deposition. The viability of the SAM barrier layer is determined by the quality of the deposited copper seed film, judged quantitatively by thin film resistivity and qualitatively by surface adhesion and morphological properties such as cracks and bubbles. Insights to the origins of problems are described and an optimal scheme described. Extensions for use as a photolithographic resist layer are suggested. Our SAM approach for TSV applications yields a 'smart' seed layer that can be used with a 'simple,' scalloped, easy to fabricate, via hole.

**I. INTRODUCTION**

Modern electronics operates at high frequencies. Inductance effects push the current in such circuits to the surfaces or edges of the conductors. This necessitates seed layers on electrodeposited conductors to have low electrical loss, as they are inherently at the edge. The electrical properties of the first few, early layers of metal on the seed are not well studies despite significant efforts in SAM-based seed layers including deposition on TaN [1, 2], and selective area deposition involving Co particles [3], sprayed-on Ag particles [4], and no seeds at all [5]. Three-dimensional (3D) integration and packaging has gained significant interest in recent years for both complementary metal-oxide semiconductor (CMOS) and radio frequency (RF) modular applications [6]. Of particular interest in the common 'via-last' packaging design is the signal performance of through-silicon vias (TSVs). The surface of the TSVs is the seed layer and hence must be low-loss. Within the next 3-5 years [7], TSV diameters ranging from 0.8-1.5 μm with aspect ratios ranging between 10:1 - 20:1 and pitch lengths between 1-4 μm are expected, meaning the seed layer makes up a larger portion of the via volume. From a materials perspective, the barrier layers [8, 9] used to minimize diffusion and electromigration effects between the conductor and substrate are also important, and are important in our solution as described below. At high frequencies (HF > 0.4 GHz), signal attenuation occurs due to inductive-capacitive (LC) delay. Because the skin depth of the conductor diminishes as a function of permeability and frequency, the current is pushed to a thin layer at the surface. Since the conduction area is limited to the penetration depth times the via perimeter, signal loss is too high if either (1) the penetration depth of the outermost conductive layer is too thin, which implies the use of a low permeability material, or (2) conductor resistivity of this outermost layer deviates significantly from its bulk value. We address the formation of very high quality 'smart' seed layers for via formation, which are crucial since the RF current flows at these outer edges.

---

[*] Corresponding Author: Hans_Hallen@ncsu.edu, 919-515-6314, fax 919-515-6538

Although specifics differ for various complex surfaces, the key aspects can all be discussed in terms of TSV creation, so we will consider them in the following discussion. TSVs are formed in three major processing steps: 1) via formation through deep reactive ion etching (DRIE), 2) sidewall deposition of insulation, diffusion barrier, and metallic seed layers, and 3) TSV metallization. Copper has long been proposed as a conductor to replace aluminum for TSV metallization due to its low resistivity and high electromigration resistance [10,11], but deposition is not without its processing difficulties. Effective copper via deposition depends on the quality of the seed layer, whose adhesion lies in creating conformal coverage of the underlying layers. High aspect ratio (>10) vias can be formed by DRIE, but without careful control of the process variables [12,13,14], possess wavy, 'scalloped' sidewalls, which inhibits conformal coverage and makes void-free metallization difficult. Copper TSV seed layers are most commonly deposited through physical vapor deposition (PVD), chemical vapor deposition (CVD), or electrochemical deposition (ECD), using oxide or polymeric based insulator layers, and barrier layers consisting of refractory metal binary and ternary compounds [15,16,17,18]. With PVD processes such as sputtering, conformal high aspect ratio step coverage is hindered as the aspect ratio increases. CVD processes such as atomic layer deposition (ALD), while conformably effective, are undesirable due to the high temperatures (> 400˚C) required for deposition and the organometallic precursors used, which can negatively affect the resistivity and diffusivity of the resulting seed layer. In recent years, electroless deposition [19,20,21,22], where the copper layer plates from solution onto a catalytically activated surface, has emerged as a viable low temperature alternative to existing TSV seed layer deposition techniques. In response, many studies [23,24,25,26,27,28], have attempted to find compatible means of electroless barrier layer deposition; however, this particular process has its drawbacks. Existing methods catalyze the underlying insulation layer, which can generate impurity diffusion into the insulation layer and silicon. Given the breadth of processing challenges, it is therefore desirable to find a TSV barrier layer suitable for electroless deposition that provides both diffusion and high temperature resistance while reducing resistive effects. The aim of this paper is thus twofold. First, we propose the use of (aminomethylaminoethyl)phenethyltrimethoxysilane (PEDA), a self-assembled monolayer (SAM) that acts as a non-metallic TSV barrier layer material, providing a conformal, catalytic surface for the electroless deposition of copper, as well as a selective patterning layer for deep UV lithography. Second, we aim to quantify the process conditions that produce high quality electroless copper seed layers upon the PEDA barrier layer, given that this is the outermost layer that determines high frequency performance. We show that it is possible to make a barrier/seed layer that is capable of rapid transition to high quality copper, all while maintaining good barrier layer characteristics. This work reviews the parts of this problem already addressed, and focuses on the remaining issues to be solved, which relate primarily to correlating the quality of the catalyzed SAM layer to the subsequent quality of the deposited copper seed layer. Briefly, we propose a 'smart seed and barrier' to work with a simple (easily fabricated) via hole, as opposed to a simply deposited seed layer in via hole that must be fabricated with tight parameter controls for uniform angle, smooth walls required by the simple seed.

**A. SAM as barrier layer**

As described earlier, a major processing challenge hindering high aspect ratio TSV Cu metallization is the deposition and performance of the barrier layer, which prevents copper diffusion into the TSV silicon. Because the seed layer requires a high temperature anneal after deposition to remove crystalline defects, an effective barrier layer must possess three major material properties: 1) high melting temperature, 2) low Cu diffusivity at elevated temperatures, 3) sufficient adhesion to both the insulation and seed layers. In addition, the skin depth limitations of RF devices require conductors with low permeability ($\mu$). Titanium, tantalum, and their nitrides [15-18] have been used as barrier layer materials, largely because of their effectiveness in aluminum ULSI metallization. Traditionally, such layers are deposited by sputtering, but deposition is hindered in high aspect ratio TSVs by the shadowing effect, which can minimize or completely block sidewall deposition. In response, electroless barrier layer deposition has been proposed for several years [23-28], with nickel and cobalt the most used base metal candidates due to their widespread industrial use and high temperature and electromigration resistance. They are alloyed with refractory metals such as rhenium, tungsten, and molybdenum to increase their high temperature resistance [23-28], along with phosphorus and boron to reduce inductive effects.

The continued use of metallic based barrier layers, however, will be hindered by two distinct design limits. In CMOS devices, it is desirable to minimize the resistive contribution of RC delay from the barrier layer, while in RF devices, it is desirable to minimize the inductive contribution of LC delay and losses from conduction within the skin depth layer due to the barrier layer. In RC limited devices, advances in scaling will require thinner barrier layers. For example, it is estimated that for 14 nm node devices, barrier layers must approach 2 nm or less [7]. However, the barrier thickness does not scale without significant diffusivity and resistance reliability issues. In LC limited devices, skin depth dependent signal propagation is constricted as frequency increases. As signal frequency increases in RF devices, the skin depth, $\delta$, of the conductor approaches the TSV diameter as scaling decreases. For example, in the SiC module described in the introduction, the maximum copper skin depth ($\delta_{max} = d_{TSV}/2$) corresponds to a usable frequency range of 8 – 30 Ghz before skin effects dominate. As skin depth decreases as frequency increases, the ability for thinner barrier layers to protect diffusion while minimally contributing inductive and resistive effects to the device will again be compromised. Ultimately, it is the material limits of existing barrier materials drives our push towards non-metallic barrier materials such as SAMs.

The advantages of a SAM barrier layer for TSVs are numerous: 1) binding groups that facilitate oxide insulator functionalization, 2), end groups suitable for catalysis, 3) selective patterning deposition, and 4) high temperature Cu diffusion protection, all at low processing temperatures (< 100˚ C). Krishnamoorthy, et al. [29] demonstrated the feasibility of the use of SAMs as barrier layers in ULSI integration processes, using bias thermal annealing. He concluded SAM layers with aromatic groups provide better barrier layer protection than aliaphatic SAMs, based on the change in leakage current density over time. Several groups [30,31,32,33] have used 3-aminopropyltrimethoxysilane (APTMS), an aliaphatic SAM layer tested in Ref. 22, as well as 3-aminopropyltriethoxysilane (APTES) [28, 34,35,36], to electrolessly deposit barrier [28,35-36] and Cu seed [30-31, 34-36] layers into TSVs with aspect ratios as large as twenty [35, 36]. Multilayers of APTMS with other materials can increase the Cu diffusion barrier [37]. Molecular dynamics studies emphasize the importance of ordered layers [38], while experimental studies note the importance of solvents, in particular the use of hydrophobic ones [39]. In addition, APTMS functionalized EL copper was recently shown to withstand copper diffusion in vacuum annealed samples ranging up to 700˚C [32]. Although there exists extensive research on the $Cu_{seed}$/APTMS and $Cu_{seed}$/APTES systems, we investigate the aromatic SAM (aminomethylaminoethyl)phenethyltrimethoxysilane (PEDA), for reasons we will describe in detail below. It contains an aromatic ring, so should act as a good barrier when the layer is dense. We monitor density by the index of refraction and thickness to insure a single layer, using contact angle and ellipsometry.

**B. SAM as patterning layer**

In our work, we ultimately seek to find a single material that facilitates both seed layer functionality and selective patterning. Photolithography using SAMs has been extensively studied and characterized [40,41,42,43,44,45,46,47,48,49]. The main advantages of SAM masks over organic photoresists are twofold. First, by careful selection of the binding and alkyl groups, such SAM layers can be selectively patterned under DUV radiation for high resolution metallization etch layers [40]. Deep UV light exposure cleaves the SAM molecule near the attachment point, removing the functional portions of the SAM later, so that the metal deposition can be selectively patterned [40]. Second, by careful selection and catalysis of the ligating end group of the SAM layer, numerous types of metallic layers, including copper, nickel, and cobalt, have been grown atop the patterned layers. [40-44, 46-49]. For our purposes, the ideal SAM layer has three characteristics: 1) a binding group that bonds to a silicon/silica surface, 2) an alkyl group that selectively absorbs DUV radiation, 3) a ligating end group that allows for electroless metallization. Prior research has focused on the use of organosilanes, specifically trimethoxysilanes, as the binding group. The methoxysilane groups react with hydroxyl groups to form siloxane bonds and liberates methanol as by-products [40-49]. These are ideal reactions for silicon, which creates high quality oxides for use for hydroxyl groups.

TABLE I.  Photolithographic characteristics of selected aromatic organosilane SAM layers from References [42] and [47]. [a]$\varepsilon_{193}$ is the molar emissivity at 193 nm in L/mol cm. [b]*Dose* is the exposure dosage at 193 nm in mJ/cm[2].

| SAM layer | $\varepsilon_{193}$[a] | *Dose*[b] | Quantum Yield |
|---|---|---|---|
| PEDA | 4.7E4 | 400 | 0.03 |
| PYR | 1.2E4 | 4500 | 0.03 |
| PTCS | 5.0E4 | 400 | 0.04 |
| CMPTS | 5.0E4 | 50 | 0.25 |

The selection of the alkyl group is guided by two principles: 1) complete DUV absorbance and alkylamine photocleavage, 2) DUV dosages and photospeeds that prevent photoablation, as opposed to photocleavage, of the underlying binding group or substrate. Initial studies [40-48] identified organosilanes with aromatic functional groups as effective SAM layers for DUV photocleavage. The phenyl groups are highly absorbing at DUV wavelengths, including the 193 nm and 248 nm lines. Studies with SAM layers aromatic groups, including phenyltrichlorosilane (PTCS) [40-42, 48], 2-(trimethoxysilyl)ethyl-2-pyridine (PYR) [43,44,46,49,50], (aminomethylaminoethyl)phenethyltrimethoxysilane (PEDA) [42, 45-47, 49], and chloromethylphenyltrichlorosilane (CMPTS) [41,42] have shown effective photocleavage of their respective aromatic groups at the Si-C bond. Our ideal SAM layer should thus possess a very short photospeed with a large quantum yield.

The dosage characteristics of the studied aromatic SAM layer are shown in Table I. Exposure doses are inversely proportional to the absorptivity of the chromophore in the film as well as the photoprocess quantum yield [47,48]. On one end of the dosage scale, PYR films, with a molar absorptivity coefficient of 1.2E4 L / mol cm at 193 nm, require a dose of 4.5 J/cm$^2$ for total Si-C photocleavage [47] which is too large for high resolution lithography. On the opposite end of the dosage scale, CMPTS films ($\varepsilon_{193\,nm}$ = 5E4 L / mol cm) have exhibited photocleavage at doses < 50 mJ/cm$^2$ at 193 nm. However, such films require an extra step by which the ligating end groups are grafted onto the remaining patterned SAM layer [42]. Both PEDA and PTCS films possess comparable molar absorptivity coefficients at 193 nm ($\varepsilon_{193}$ = 4.7E4 L/mol cm vs. 5E4 L/mol cm, respectively), resulting in comparable photocleavage dosages of 400 mJ/cm$^2$. However, since PTCS does not include the extra ligating end group, it does not meet the necessary characteristics for a suitable SAM mask layer, leaving PEDA as the best SAM layer choice for both patterning and etch layer protection due to its DUV absorbing aromatic alkyl group and its diamine ligating end group [47].

**C. SAM catalysis for copper seed layer**

Since prior groups have demonstrated the high temperature resistance [32] and patterning selectivity [40, 49] of PEDA barrier layers for use in Cu metallization, we aim in this work to correlate the catalytic behavior of the activated PEDA layers to the eventual as-deposited quality of the copper seed layer. Surface catalysis is accomplished by treating the substrate with a colloidal suspension of Pd and Sn species. The colloidal particles consist of a Pd metallic core, 5 to 20 nm in diameter, surrounded by an Sn-rich layer. The central core is comprised of a low-valence, Pd rich intermetallic species, which acts as the actual catalyst in the initial reduction process leading to plating. Prior research [44] has found that the Pd/Sn catalyst adheres sufficiently to enable electroless metallization of surfaces modified with terminal olefin, amine, thiol, hydrocarbon, heterocyclic, phosphocholine, and other functional groups. The Pd/Sn catalyst, however, does not adhere to surfaces with high levels of free Si-OH groups. The diamine group found in PEDA is thus suitable for Pd/Sn catalysis. The critical characteristics of a successful catalysis layer are small particle size, small particle size distribution, and complete particle coverage of the SAM layer. If the particle size (~10-20 nm) or distribution ( ± 20 nm) is too large, the grain size of the resulting EL Cu will be uneven, as larger catalyst particles form larger Cu grains, and minimize the contributions of smaller particles. Grain size considerations become even more critical for TSV seed layers as the aspect ratio of the catalyzed surfaces increase. Initial TSV SAM barrier/seed studies [28, 30-35] catalyzed their SAM layers using a gold based colloidial suspension, due to the inability to control the particle size distribution of traditional Pd/Sn based catalysis. Although successful in creating monodisperse suspensions, Au diffusivity into silicon is undesirable. Several groups [44,46,49] have previously addressed catalysis issues in EL nickel metallization using careful Pd particle size control and Sn layer elimination, and whose strategies are employed in our EL copper metallization. Our catalysis method, as described by Brandow, et al. [49] creates an initial rapid nucleation of Pd particles in solution, which is subsequently neutralized to sufficiently control growth. In such a manner both particle size and size distribution can be controlled, leaving only catalysis time to be optimized.

## II. EXPERIMENT

Our copper metallization technique is based on an approach described for nickel deposition [37, 39, 42]. To characterize the metallization process we use the sessile drop method [40, 43-48] to determine optimal PEDA deposition time. Once the copper layers are deposited, we determine the copper quality by measuring the copper thickness and film resistivity as a function of catalyst and copper bath deposition time. Deionized water with > 18 MΩ resistivity was used throughout the experiment. PEDA used in the experiment was obtained from Gelest, Inc. All solvents were obtained from Fisher Scientific. Disodium tetrachloride palladate, $Na_2PdCl_4 \cdot 3H_2O$, was obtained from Aldrich Chemical Company. All other acids were obtained from Fisher Scientific. The copper bath is a commercially available solution obtained from Transene, Inc. The PEDA samples were prepared on native oxide Si wafers (n-type (100), Wafer World). Two sets of six 1" square silicon samples were cleaned in a piranha etch consisting of a (1:1 v/v) $H_2SO_4/H_2O_2$ solution for one hour, rinsed in DI water, then passivated in a 3% HF solution for 7 minutes. After another DI rinse, the first set of samples was immersed in a 1% PEDA solution for 15 minutes, then submerged in a colloidal palladium catalyst dispersion for an additional 15 minutes, following the recipe of Brandow et al. [49]. Their recipe called for a filtering of the final solution in a .22 μm Teflon filter; in lieu of the filter, we used a Stokes' Law analysis to estimate the time for particles of this size or greater to float down to the bottom of our container, keeping it refrigerated at 8°C to hasten excess growth of particles of the desired size. The samples were then rinsed in isopropanol for 2 minutes, submerged in the copper solution at 40°C, and removed in sequence after 15, 30, 60, 120, 240, and 480 seconds of deposition time. The second set of samples was submerged in PEDA for 15 minutes, then dipped in a separate Pd solution, removed in sequence after 1, 2, 4, 8, 15, and 30 minutes, rinsed in isopropanol for two minutes, and finally submerged in a separate copper bath for 2 minutes. The contact angles of the PEDA films were measured using a custom designed sessile drop system. The thicknesses of the copper films were measured using a Dektak profilometer, and resistivities measured with a Keithley four point probe.

## III. RESULTS AND DISCUSSION

Surface functionalization of PEDA requires a hydrogen-passivated surface prior to deposition. If the substrate is not properly hydrolyzed, the quality of the PEDA deposition will be compromised. We therefore establish the optimal PEDA deposition time by measuring the sessile drop contact angle as a function of PEDA deposition time at a fixed passivation time of 7 minutes, as stated in the Experiment section. Figure 1 shows contact angle measurements after PEDA deposition as a function of deposition time. According to prior studies [45], the contact angle for a high quality PEDA layer is approximately 52 – 55°. Figure 1 shows that after an initial increase in contact angle during the first 15 minutes of deposition, the contact angle eventually settles into the 52-55° range, suggesting the optimal PEDA deposition time is between 15 to 30 minutes. This provides a high quality single layer with minimal possibility of multiple layer growth.

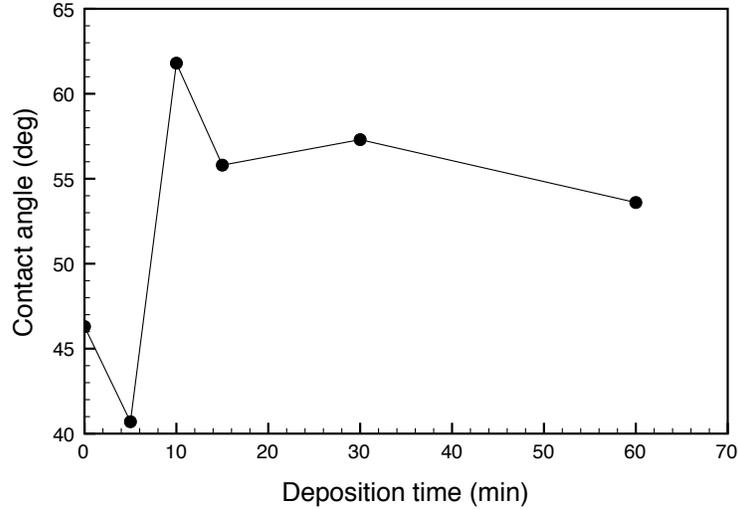

FIG. 1. Contact angle data as a function of PEDA deposition time.

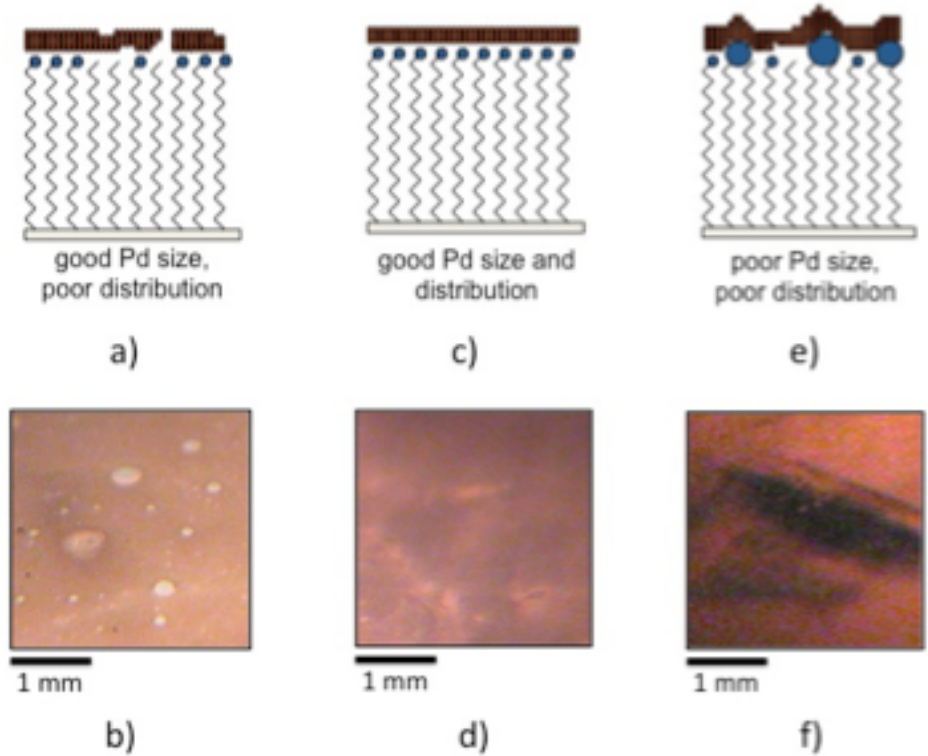

FIG. 2. Schematic and images of two minute electroless copper growth on silicon showing two failure modes and the optimal case. The quality of the copper film growth is determinant on three factors: 1) PEDA distribution (wavy lines) on the silicon substrate, 2) PD2 (blue dots) catalyst size and distribution, 3) copper (smaller dots) bath deposition time. (a,b) Two minute PD2 catalysis, resulting in low Pd distribution and copper films with voids. (c,d) Eight minute PD2 catalysis, resulting in good Pd distribution and high quality copper films. (e,f) 30 minute PD2 catalysis, resulting in excess growth of Pd particles and rough, poorly adhering copper film.

The next step in the deposition process is PEDA catalysis, and its role in the quality of the copper seed layer. The ideal PD2 conditions should allow for the deposition of a reproducibly thin (~ 100 Å) seed layer, whose sheet resistivity, $\rho_s$, should approach that of bulk copper, approximately 0.0168 $\mu\Omega$-m [50]. To determine the dependence of copper thickness and resistivity on growth variables, we establish a reference point in an independent variable space that includes four degrees of freedom: PEDA deposition time, PD2 catalysis deposition time, copper bath temperature, and copper bath deposition time. As described above, the optimal PEDA deposition time is ~ 15 minutes, eliminating one degree of freedom. The bath temperature is kept constant at 40° C, per the deposition instructions of the Transene solution, thus eliminating another degree of freedom. This leaves two parameters, PD2 and copper bath deposition time, to adjust, while noting the PD2 and copper film thickness and optimizing the resistivity and film mechanical properties: morphology and substrate adherence. After determining the thickness growth rate, the resistivity is easily quantified and we optimize it by independently varying the parameters to get 'line-cuts' of the surface of the resistivity as a function of these two parameters. These line-cuts cross at a reference point, chosen not too far from the optimal conditions. The growth conditions of our chosen reference point is 15 minutes of PEDA growth, 15 minutes of PD2 growth and 2 minutes of copper growth. For catalysis, there is an ideal balance between both the particle size and palladium ion concentration collected by each amine end group of PEDA. Figure 2 shows in the top row a schematic of the PD2 development in relation to its resultant copper film growth quality in the bottom row. The proper catalyst layer is a uniformly sized, tightly distributed Pd catalyst layer that yields high quality smooth films that adhere well to the substrate, as shown in Figure 2c,d). At one processing extreme, too short a PD2 deposition time results in too few Pd ions collected per amine group, resulting in broken, non-uniform copper layers. These layers do not adhere well to the PEDA and tend to crack, bubble and peel, as shown in Figure 2a,b). At the other processing extreme, too long a PD2 deposition creates Pd particles ions of varying size per amine group, resulting in a rough copper layer, sporadically attached to the underlying PEDA layer, that adheres poorly to the silicon substrate, as shown in Figure 2e,f).

The film thickness and resistivity data as a function of PD2 and copper deposition time is shown in Figures 3a) and 3b). For a fixed PD2 deposition time of 15 minutes, there is a linear dependence of copper thickness to copper bath deposition time, ranging from 0.4 nm at 30 seconds of deposition time to 600 nm at 8 minutes of deposition time, corresponding to a deposition rate of approximately 0.184 mil/hr. This is close to the nominal deposition rate of 0.2 mil/hr. For a fixed copper deposition time of 2 minutes, the measured copper thickness is initially large at short PD2 deposition times (e.g., 345 nm at 1 minute of PD2 deposition) due to direct Cu-PEDA bonding. As PD2 deposition time increases however, copper thickness decreases due to direct Cu-PD2 bonding, reaching a minimum of 18.2 nm at 15 minutes of PD2 deposition before rising to a value of 228 nm at 30 minutes as the PD2 particle size distribution increases.

Copper resistivity also reflects film quality changes with Pd and Cu growth times. For the copper bath time variation, short bath times (i.e. < 60 sec), yield resistivities approaching that of bulk copper. The resistivity spikes up to 6.92 µΩ-m at the 1 minute mark, as the relatively thick (60 nm) copper layer grown atop the Cu-PEDA bond begins to relieve stress and form defects. As time progresses, the resistivity then drops to a minimum of 0.0415 µΩ-m at 4 minutes, still an order of magnitude higher than that of sheet copper resistivity, due to the growth inconsistencies of the underlying 15 minute PD2 layer. The resistivity dependence on PD2 deposition time surface trends in a similar fashion, with a value of 0.5 μΩ-m at 1 minute of PD2 deposition, due to Cu-PEDA bonding and other film defects. Bulk copper resistivity is approached near 4 minutes, reaching a minimum of 0.00936 μΩ-m at 8 minutes. The resistivity rises for longer catalyst deposition times. This dependence on growth reflects the stages of PD2 development outlined in Figure 3.

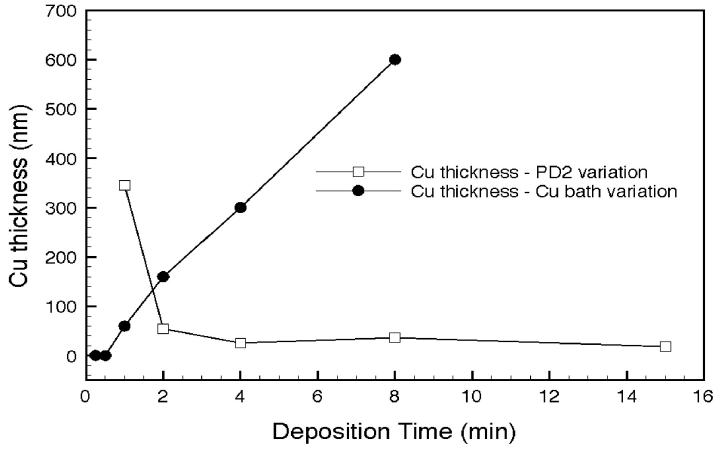

a)

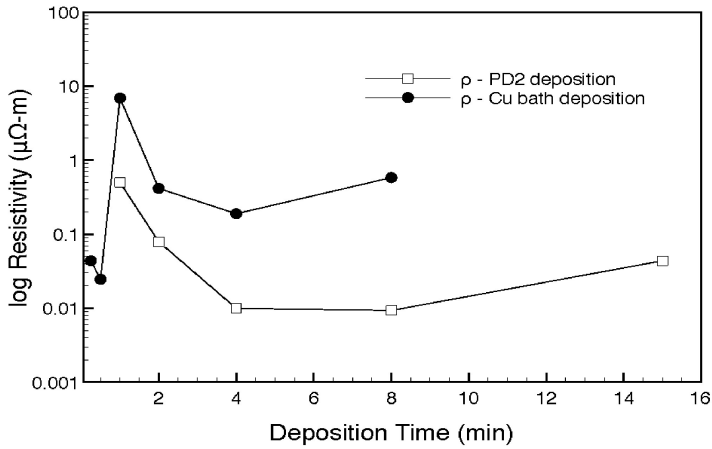

b)

FIG. 3. Copper layer characteristics. The squares represent variations in the PD2 catalyst deposition time for a fixed 2 minute copper deposition time, and the circles represent variations in the copper bath deposition time for a fixed 15 minute PD2 catalyst deposition time. a) Copper layer thickness for a fixed PEDA deposition time of 15 minutes. b) Copper layer resistivity for a fixed PEDA deposition time of 15 minutes.

   Comparisons of the data in Figure 3 suggest that the optimal growth conditions are 15 minutes of PEDA deposition, 4 minutes of PD2 deposition, and 4 minutes of Cu deposition at 40°C, since the resistivity data for both parameter variations trend toward the value of bulk copper sheet resistivity. The four minutes of PD2 deposition differs from the reference value of 15 minutes, as it gives a marked improvement. The copper deposition time of four minutes rather than two is less significant. The broad minimum against variation of each parameter suggests that these values are robust, and slightly better results would be attained near the minimum of these curves at optimal growth conditions. An important point is that a resistivity near the bulk value is reached at thicknesses below 100 nm. As stated before, lower loss at RF is attained when the transition to high quality copper is fast. The size and distribution of Pd catalyst used here our $Cu_{seed}$/PEDA system seems to outperform the APTMS and APTES systems used in other work [30-36], whose subsequent Cu layer resistivity is on average 2-3 times higher than the bulk Cu resistivity. Our system also incorporates the benzene ring within the PEDA molecule, lacking in the aliphatic SAMs used by the other groups. The aromatic ring has been shown [29] to improve both thermal stability and reduce diffusion (likely due to an increase in the layer density at the ring), hence improving its action as a barrier layer.

## IV. SUMMARY AND CONCLUSIONS

We have identified a barrier/seed system to seed complex, non-line of sight surfaces with a copper layer before further electrochemical growth to fill vias or smaller traces. At high frequencies, the conductivity of this seed layer is important, so it is important that its characteristics be optimized. We considered the variation of four parameters: catalysis deposition time, copper deposition time, film resistivity, and film thickness, and used film structure, adherence, and resistivity to find optimal parameter values. The base PEDA layer is straightforward to optimize, while the Pd catalyst layer strongly and qualitatively impacts both structural and electrical properties of the copper. We qualitatively and quantitatively evaluate these properties.

It is vital that the PEDA and catalyst layers are deposited as uniformly as possible. With a dense PEDA and even, well-distributed PD2 layers, too short a copper deposition time does not allow enough time for uniform growth across a uniform PD2 layer, but too long a deposition time creates copper overgrowth, with poor substrate adhesion. Optimal copper resistivity is reached between 4-8 minutes of deposition growth, corresponding to a thickness of 25-35 nm, reached approximately after 2 minutes of deposition growth. We are confident that the uniformity of these electrolessly deposited copper layers is sufficient to meet the demands of 3D packaging applications – a 'smart' seed layer for 'simple' via holes, and other complex surface metallization needs.

## ACKNOWLEDGEMENTS

This work was supported by AFRL-WPAFB (Grant FA8650-04-2-1619).

9 Z. Huang, Y. Zhou, and W. He, "A combination of self-assembled monolayer and hydrophobic conformal coating for anti-corrosion of Cu/NiP/Au 3D circuitry in artificial sweat solution," *Surf. Coat. Technol.* (Netherlands), **320**:126 – 31, (2017).
10 J. Sun, K. Kondo, T. Okamura, S. Oh, M. Tomisaka, H. Yonemura, M. Hoshino, and K. Takahashi, J. Electrochem Soc., **150**, G355 (2003).
11 K. Kondo, T. Yonezawa, D. Mikami, T. Okubo, Y. Taguchi, K. Takahashi, and D. P. Barkey, J. Vac. Sci. Technol. B. **152**, H173 (2005).
12 F. Laermer and A. Urban, Microelectronic Engineering, **67-68**, 349, (2003).
13 S. C. Chen, Y.C. Lin, J. C. Wu, L. Horng, and C. H. Cheng, Microsyst. Technol., **13**, 465, (2007).
14 X. Wang, W. Zeng, G. Lu, O.L. Russo, E. Eisenbraun, J. Vac. Sci. Technol. B. **25**, 1376 (2007).
15 K. C. Park, K. B. Kim, I. Raajimakers, and K. Ngan, J. Appl. Phys., **80**, 5674 (1996).
16 T. Nguyen, L. J. Charneski, and D. R. Evans, J. Electrochem Soc., **144**, 3634 (1997).
17 M. H. Tsai, S. C. Sun, C. E. Tsai, S. H. Chuang, and H. T. Chiu, J. Appl. Phys., **79**, 6932 (1996).
18 S. M. Rossangel, J. Vac. Sci. Technol. B. **20**, 2328 (2002).
19 V. Dubin, Y. Shacham-Diamand, B. Zhao, P.K. Vasudev, and C. H. Ting, J. Electrochem Soc., **144**, 898 (1997).
20 Y. Shacham-Diamand, S. Lopatin, Electrochimica Acta, **44**, 3639 (1999).
21 H. Hsu, K. Lin, S. Lin, and J. Yeh, J. Electrochem Soc., **148**, C47 (2001).
22 F. Santagata, C. Farricelo, G. Fiorentino, H. W. van Zeijl, C. Silvestri, G. Q. Zhang, and P. M. Sarro, J. Micromech. Microeng., **23**, 055014 (2013).
23 A. Kohn, M. Eizenberg, Y. Shacham-Daimand, Y. Swerdlov, Mater. Sci. Eng. A Soc., **302**, 18 (2001).
24 T. Osaka, N. Takano, T. Kurokawa, T. Kaneko, K. Ueno, J. Electrochem. Soc., **149**, C573 (2002).
25 D. Liu, Z. Yang, C. Zhang, Materials Science and Engineering B **67**, 623 (2005).
26 H. Einati, V. Bogush, Y. Sverdlov, Y. Rosenberg, Y. Shacham-Diamand, Microelectronic Engineering, **82**, 623 (2005).
27 S. B. Antonelli, T. L. Allen, D. C. Johnson, and V. M. Dubin, J. Electrochem. Soc., **152**, J120 (2005).
28 M. Yoshino, Y. Nonaka, J. Sosano, I. Matsuda, Y. Shacham-Diamand, Electrochimica Acta, **51**, 916 (2005).
29 A. Krishnamoorthy, K. Chanda, S. P. Muraka, G. Ramanath, and J. G. Ryan, Appl. Phys. Lett., **78**, 2467 (2001).
30 E. Glickman, A. Inberg, N. Fishelson, and Y. Shacham-Diamand, Microelectronic Engineering, **84**, 2466 (2007).
31 T. Asher, A. Inberg, E. Glickman, N. Fishelson, and Y. Shacham-Diamand, Electrochimica Acta, **54**, 6053 (2009).
32 S. Sharma, M. Kumar, S. Rani, and D. Kumar, "Deposition and Characterization of 3-Aminopropyltrimethoxysilane Monolayer Diffusion Barrier for Copper Metallization," *Metall. Mater. Trans. B Process Metall. Mater. Process. Sci.*, 2014.
33 S. Sharma, M. Kumar, S. Rami, A. Singh, B. Prasad, D. Kumar, AIP Conference Proceedings, **1536**, 1163 (2013).